\documentclass[useAMS,usenatbib]{mn2e}
\usepackage{amsmath}
\usepackage{graphicx}
\usepackage{natbib}
\usepackage{epstopdf}
\usepackage{float}
\usepackage{fixltx2e}

\begin{document}

\title[Magnetic field expulsion from NS
cores]{Simulated magnetic field expulsion in neutron star cores}

\author[J. G. Elfritz et al.]{J. G. Elfritz,$^{1}$\thanks{E-mail:
j.g.elfritz@uva.nl} J. A. Pons,$^{2}$ N. Rea,$^{1,3}$ K.
Glampedakis,$^{4,5}$ and D. Vigan\`{o}$^{3}$\\
$^{1}$Anton Pannekoek Institute for Astronomy, University of
Amsterdam, Postbus 94249, 1090GE Amsterdam, The Netherlands\\
$^{2}$Departament de F\'{i}sica Aplicada, Universitat d'Alacant, Ap.
Correus 99, E-03080 Alacant, Spain\\
$^{3}$Institute of Space Sciences (ICE, CSIC-IEEC), Campus UAB, Carrer
de Can Magrans s/n, E-08193 Barcelona, Spain\\
$^{4}$Departamento de Fisica, Universidad de Murcia, E-30100 Murcia, Spain\\
$^{5}$Theoretical Astrophysics, University of T\"{u}bingen, Auf der Morgenstelle 10, D-72076, T\"{u}bingen Germany}

\date{\today}

\pagerange{\pageref{firstpage}--\pageref{lastpage}} \pubyear{2015}

\maketitle

\label{firstpage}

\begin{abstract}

\noindent The study of long-term evolution of neutron star (NS) magnetic fields is key to understanding the rich diversity of NS observations, and to unifying their nature despite the different emission mechanisms and observed properties. Such studies in principle permit a deeper understanding of the most important parameters driving their apparent variety, e.g. radio pulsars, magnetars, x-ray dim isolated neutron stars, gamma-ray pulsars. We describe, for the first time, the results from self-consistent magneto-thermal simulations considering not only the effects of the Hall-driven field dissipation in the crust, but adding a complete set of proposed driving forces in a superconducting core. We emphasize how each of these core-field processes drive magnetic evolution and affect observables, and show that when all forces are considered together in vectorial form, the net expulsion of core magnetic flux is negligible, and will have no observable effect in the crust (consequently in the observed surface emission) on megayear time-scales. Our new simulations suggest that strong magnetic fields in NS cores (and the signatures on the NS surface) will persist long after the crustal magnetic field has evolved and decayed, due to the weak combined effects of dissipation and expulsion in the stellar core.

\end{abstract}


\section{Introduction}\label{section:introduction}

Many of the aspects concerning the internal structure and long-term evolution of neutron star (NS) cores, and what core-driven signatures in the NS emission may be detectable, are still largely unknown and subject of dedicated studies. These open questions span the most extreme regimes of nuclear and plasma physics, environments which are impossible to reconstruct terrestrially. The origin of the large scale proto-NS magnetic field, soon after the supernova explosion, may possibly control the initial NS morphology throughout the hot interior. Soon after formation, the neutrino emission quickly drives the internal temperatures down to $10^{9}-10^{10}$\,K, with a quasi-stationary internal temperature distribution \citep{Page2004,Yakovlev2004}. The remaining global magnetic field description may be regular due to fast self-relaxation \citep{Flowers1977}, or otherwise pseudo-ordered with intrinsic linkage between poloidal and toroidal components \citep{Braithwaite2006}. Once the crustal ion lattice crystallizes within a few days or weeks after birth, the Hall-driven magnetic field dynamics in the crust irreversibly constrain the stellar magneto-thermal-rotational evolution according to state-of-the-art numerical simulations (Vigan\`{o}, Pons \& Miralles 2012; Vigan\`{o} et al. 2013a; Gourgouliatos \& Cumming 2015a).

Anomalous X-ray pulsars (AXPs) and Soft Gamma Repeaters (SGRs), collectively called magnetars, sustain dipolar magnetic fields of $B_{\mathrm{dip}}\sim10^{14}-10^{15}$\,G, as inferred by assuming these sources to be rotating magnetic dipoles (see \cite{Mereghetti2008} for a recent review). However, the true dipolar magnetic fields of AXPs and SGRs could also be influenced by highly twisted magnetar magnetospheres \citep{Beloborodov2009}, where only now the first physics-based numerical models are coming to full fruition \citep{Philippov2014,Chen2014}. There is currently no consensus over the origins of such ultra-magnetic stars. Especially massive or magnetic progenitors could give rise to such strong internal fields \citep{Spruit2008}, potentially in conjunction with $\alpha-\Omega$ dynamo \citep{Duncan1992}. Magneto-rotational instabilities \citep{Guilet2015}, Tayler-Spruit dynamo \citep{Tayler1973, Spruit1999}, and post-infall convection (Obergaulinger, Janka \& Aloy 2014) could also further amplify the progenitor magnetic field. Even after the formation of the solid crust, magnetar flaring activity and super-strong fields can be driven by density-shear instability in the NS crust \citep{Gourgo2015b}, explaining the localization of such active magnetic regions.

\cite{Vigano2013} have recently performed a comprehensive study of the magnetic and thermal evolution of isolated NSs, exploring the influence of their initial magnetic field strength and geometry, their mass, envelope composition and relevant microphysical parameters. The careful inclusion of microphysics activation, based on the most current literature (Potekhin \& Chabrier 2013; Potekhin, Pons \& Page 2015), has proven especially critical for computing standard cooling curves for magnetized neutron stars. This work was the first numerical model to successfully handle such large magnetic field strengths ($>$\,10$^{14}$\,G) while advancing the Hall term utilizing a novel, stable field-advance algorithm throughout the first megayears (Myrs) of stellar evolution. Using this numerical model, Pons, Vigan\`{o} \& Rea (2013) showed that a highly resistive layer in the deep crust is a crucial ingredient for enhancing dissipation of magnetic energy of high-field NSs. In complementary studies, \cite{Gourgo2013} and \cite{Gourgo2014} explored long-term magnetic evolution using their own Grad-Shafranov-based recipe, specifically designed to probe the NS evolution as the Hall drift saturates, and thus as the global field approaches a Hall equilibrium. They revealed that isolated NS magnetic evolution consists of three basic phases, and that the internal fields relax to a so-called Hall attractor state, a process that is largely insensitive to the initial conditions. Recent 3D simulations suggest that the Hall-induced small scale magnetic features persist in the NS crust on longer time scales than in axisymmetric 2D simulations, although the global evolution still tends to the dipolar Hall attractor \citep{Wood2015}. It is not immediately clear whether self-consistent thermal evolution drives deviations from such an attractor in the late evolution; the aforementioned models use time-independent phenomenological descriptions for the electrical conductivity, instead of solving the coupled magnetic and thermal evolutionary equations (see \S\ref{section:MToverview}). We leave such a study for future work.

One limitation of current evolutionary modeling has been that, previously, the NS core was assumed to be non-interacting. Expulsion of magnetic field from the core could potentially influence the crustal evolution and thus observable broadband emission, and this has not yet been taken into account by numerical simulations. If the local magnetic field strength is below the upper critical value of $\sim 10^{16}$\,G, regions of the core are expected to exhibit type-II superconductivity; in this case the magnetic field is quantized into an array of so-called fluxoids. This has implications for the internal magnetic evolution and for the resulting observational signatures. A variety of physical mechanisms have been proposed to account for the expulsion of flux from the NS core \citep{Baym1975,Muslimov1985,Jones1987,Goldreich1992,Konenkov2000}, and no popular consensus within the research field has been reached regarding the mechanism (or combination thereof) ultimately responsible.

In this study we focus attention on the magnetic field evolution of the core and crust of isolated NSs in regimes relevant to magnetars as well as the moderate-field pulsars. We discuss the prominent theoretical processes considered to be potentially responsible for core flux expulsion, and we perform self-consistent magneto-thermal numerical simulations in two spatial dimensions which account for these mechanisms in a complete manner. Specifically, our simulations account for four effects: (a) magnetic buoyancy due to pressure imbalance within quantized magnetic fluxtubes, (b) viscous drag due to scattering between electrons and the magnetic field, (c) the Magnus force, and (d) forcing that results from the release of fluxoid tension (Muslimov \& Tsygan 1985; Glampedakis, Andersson \& Samuelsson 2011a). These are each discussed in detail in \S\ref{section:allforces}.

We acknowledge that our numerical model omits important physics in the core. We ignore the effects of simple, uniform rotation for two fundamental reasons: firstly, highly magnetized NSs spin down very efficiently and reach long spin periods quickly; secondly, the number of neutron vortices $N_v$ is many orders of magnitude less than the number of magnetic fluxoids $N_f$, thus one expects a very weak feedback in the magnetic field evolution from simple rotational evolution, since the mutual friction contribution term scales as $N_v/N_f$. In addition, we are interested in the study of the first $\sim 10^6$ yrs of a neutron star, when it is still bright enough for its thermal emission to be directly observed. The time-scale for ambipolar diffusion, assuming normal matter (non-superfluid and non-superconducting), is a few hundred kiloyears for $B \approx 10^{14}$\,G \citep{Goldreich1992}, and becomes even longer if we admit superfluid matter in the NS (Glampedakis, Jones \& Samuelsson 2011b). Therefore, we leave for future work the study of the effects of simple rotation and ambipolar diffusion resulting from inter-species drag forces. 

Another caveat, probably the most critical one, is that our formulation does not self-consistently include hydrodynamical evolution of constituent species in the core. The hydrodynamics couple to the rotational evolution, where in the superfluid limit a number of unique consequences could emerge. Differentially rotating superfluids are susceptible to torque oscillations \citep{Peralta2005, Melatos2007}, which are likely an ingredient in explaining timing irregularities such as glitches. Pinning of superfluid vortices to the proton-electron plasma in the core also excites hydromagnetic modes \citep{vanHoven2008} and instabilities (Glaberson, Johnson \& Ostermeier 1974, Link 2012a). Pinning of vortices to the nuclear lattice in the inner NS crust admits the excitation of additional unstable modes \citep{Link2012crust}. Thus in both the inner crust and outer core, strongly nonlinear hydrodynamically-driven effects probably impact the evolution of the NS magnetic field. These fundamental modes can drive the fluxoid array into a tangled state in the outer core and inner crust in the low-field limit, permitting a forward cascade to fully developed superfluid turbulence (Andersson, Sidery \& Comer 2007; Link 2012a,b). While we acknowledge that the hydrodynamical evolution in (at least) the outer NS core could be the predominant evolutionary driver, coupling such physics to the global magneto-thermal evolution is beyond the scope of this work. In addition, these short time-scale dynamical effects rely on rotation, a process not relevant within our numerical framework. Thus we would expect bulk hydrodynamical flows to be the most significant feature on megayear time-scales, and acknowledge our omission thereof. Finally, we also neglect direct interactions between individual fluxtubes, while recognizing that densely packed configurations may be relevant for magnetar interiors. 

The paper is structured as follows. An overview of the numerical model is provided in \S\ref{section:MTmodel}, with a focus on the quantitative description and derivation of the dynamics of the constituent species in the core, as well as the core magnetic field. Analytic estimates of flux expulsion time-scales are explored in \S\ref{section:analyticestimates}. Simulation results are presented in \S\ref{section:simulationresults} for a representative selection of initial magnetic field prescriptions.  \S\ref{section:discussion} contains a discussion of these results in the context of prior work, of current observations, and how these results may assist in constraining prevailing theories.

\section{Description of the 2D numerical simulations: physical processes and implementations}\label{section:MTmodel}

\subsection{Basic overview of the magneto-thermal code}\label{section:MToverview}

In this study we focus on the evolution of the magnetic and thermal properties of a simulated neutron star during the first megayear of the NS life cycle. Extensive descriptions of the numerical code used to perform the simulations may be found elsewhere \citep{Vigano2012}, but a brief synopsis of its maturity is in order. 

The first numerical studies of the Hall effect on dissipation of magnetic fields in the crust were presented in \cite{Hollerbach2002, Hollerbach2004}. \cite{Pons2007} used a spectrally decomposed field description, combined with a phenomenological model to mimic the thermal evolution and thus isolate their study, to focus on the magnetic evolution alone. Aguilera, Pons \& Miralles (2008) later outlined a detailed complementary investigation using the converse approach. There, all neutrino emission mechanisms were included, alongside activation of superfluidity and superconductivity, and rigorous computation of thermal conductivities, as well as all other necessary ingredients. Pons, Miralles \& Geppert (2009) coupled these two separate 2D models and performed the first self-consistent numerical simulations of coupled magnetic and thermal evolution in NSs, emphasizing the importance of such coupling for realistically representing NS evolution over megayear time-scales. A succession of follow-up studies (see \cite{ViganoThesis} and references therein) further refined earlier works by constructing a more robust field advance algorithm and uncovering observational signatures of high-field sources that were not possible with other existing numerical codes. 

Our current model computes the evolution on a static, staggered, multi-scale grid in radius and polar angle ($r$ and $\theta$ in spherical coordinates) throughout the stellar core and crust under the assumption of symmetry in the azimuthal direction, i.e. $\partial/\partial\phi\rightarrow 0$. We use a magnetic field grid of 100 $\times$ 100 cells in $r$ and $\theta$. The radial grid in the core is relatively coarse (40 radial points), while increased resolution is used in the crust (60 points) where the Hall term in the magnetic induction equation dominates the evolution. Because the Hall drift can result in highly localized structures in the magnetic field topology, and thus strong current sheets, higher spatial resolution is required in the crust. There are two fundamentally coupled equations that the model iteratively solves: the thermal evolution equation and the magnetic induction equation:\footnote{The code includes relativistic red-shift corrections for a spherically symmetric metric, but for clarity we omit these factors and refer the interested reader to \cite{Vigano2013} for details.}

\begin{equation}
c_v \partial_t T-\vec{\nabla}\cdot\left[\hat{\kappa}\cdot\vec{\nabla} T \right]=\sum_i Q_i ,
\label{equation:MTtemperature}
\end{equation}
\begin{equation}
\partial_t\vec{B}=-c\vec{\nabla}\times \vec{E} .
\label{equation:MTinduction}
\end{equation}

\noindent In equation (\ref{equation:MTtemperature}), $T=T(\vec{r},t)$ is the local thermodynamic temperature, $c_v(\vec{r},t)$ is the specific heat capacity, and $\hat{\kappa}(\vec{r},t)$ is the thermal conductivity tensor, which is generally anisotropic in the presence of magnetic fields. The $Q_i$ terms account for local sources and sinks of thermal energy per unit volume, per unit time. Typically these terms are the Joule heating rate $+Q_j$ and the neutrino emission rate $-Q_{\nu}$, but may also include e.g. accretion-induced heating. The computations required to track the thermal evolution are based on the cooling code described by \cite{Aguilera2008} and references therein; the microphysically-motivated computations of $c_v$, $\hat{\kappa}$ and $Q_{\nu}$ are not trivial, but are well described in the literature (also see Ch. 4 of \cite{ViganoThesis}).

In parallel with the temperature evolution equation, we advance the magnetic induction equation (\ref{equation:MTinduction}). The explicit form of the electric field $\vec{E}$ depends on the physical assumptions and the local description of matter, and is different in the liquid core and solid crust. In the NS crust, charged components constitute normal, non-superconducting matter and we assume that there is no fluid motion (we neglect the mass flow carried by the slow electron current). The corresponding crustal electric field is 

\begin{equation}
c \vec{E} = \eta\vec{\nabla}\times \vec{B} + f_H\left(\vec{\nabla}\times \vec{B} \right)\times\vec{B}.
\label{equation:crustinduction}
\end{equation}

\noindent where the first term on the right side of equation (\ref{equation:crustinduction}) accounts for Ohmic dissipation, and the second term is the Hall drift. The Hall term is non-dissipative, but influences the configuration of currents, mediates the transfer of magnetic energy between poloidal and toroidal magnetic field components, and can drive dynamics down to smaller spatial scales during the evolution. The Hall pre-factor is $f_H(r)$=$c/4\pi e n_e(r)$, and $\eta(\vec{r},t)$=$c^2/4\pi\sigma(\vec{r},t)$ is the magnetic diffusivity. The electrical conductivity $\sigma(\vec{r},t)$ is a strong function of local temperature (see figure 1 of \cite{Pons2009}), which therefore couples the Ohmic dissipative efficiency to the thermal evolution. In our model, the crustal conductivity is computed from the public code developed and maintained by A. Potekhin (1999)\footnote{\texttt{www.ioffe.rssi.ru/astro/conduct/condmag.html}.}

The dynamical evolution of the NS core is much more complex, and is poorly understood. In principle the magnetic field should be computed self-consistently alongside the multi-fluid hydrodynamical evolution, but such a complete treatment is far beyond the reach of existing numerical models. If the core is considered as normal matter, as in the crust, then Ohmic diffusion controls the slow decay of the magnetic field. Electrical conductivities are predicted to be many orders of magnitude greater than in the crust, implying dissipation timescales of $\sim 10^7$\,yrs. Figure (\ref{fig:eosparams}) shows a radial profile of the electrical resistivity (inverse conductivity) under the assumption of normal matter in the core, at high temperature (young NS) calculated from the models of \cite{Gnedin1995} and Baiko, Haensel \& Yakovlev (2001). If in the NS core the protons constitute a superconducting fluid, then the electric field is no longer purely Ohmic, but follows from the electron momentum equation. \citet{Glampedakis2011a} have shown that the type-II superconducting electric field appropriate for the NS core is

\begin{equation}
\vec{E}=-\frac{1}{c}\vec{v}_e \times\vec{B}-\frac{1}{e}\vec{\nabla}\left(\mu_e+m
_e\Phi\right)-\frac{1}{e n_e}\left(\vec{F}_{ne}+\vec{F}_{pe}\right)
\label{equation:coreEfield1}
\end{equation}

\noindent after dropping the small inertial terms. Here the mutual friction terms $\vec{F}_{xy}$ arise from the multi-fluid MHD formalism. We do not advance hydrodynamic equations for any species present in the core, but rather include mesoscopically averaged dynamics which enter via the $F_{xy}$ terms, most importantly the proton-electron coupling term $F_{pe}$, because $F_{ne} \ll F_{pe}$. The mesoscopic averages allow us to approximate the large-scale drift of the magnetic field in the core, and thus track its evolution. In \S\ref{section:allforces} we discuss in more detail the relation between electron fluid motion and mesoscopic fluxtube motion, and begin to construct the core induction equation by considering the terms which contribute to the total mutual friction. In \S\ref{section:fullforcebalance}, we derive the final form of the core induction equation, and summarize the numerical procedure.

For the simulations discussed in this manuscript, we use the canonical NS mass of 1.4 M$_{\odot}$, stellar radius of 11.5\,km, and a core radius of 10.8\,km. The NS structure is determined from a Skyrme-type equation of state (EOS) to describe the liquid core and the solid crust; this is based on the SLy nuclear interaction \citep{Douchin2001}. We also employ the classic BPS relation in regions where the density is below the neutron drip point (Baym, Bethe \& Pethick 1971). We use the same energy gap models of Ho, Glampedakis \& Andersson (2012) for neutrons in the core (deep triplet $^3$P$_2$), neutrons in the crust (singlet $^1$S$_0$), and protons in the core (singlet $^1$S$_0$). All simulations take a modest value for the impurity parameter in the deep inner crust, $Q_{\mathrm{imp}}=5$ (see \cite{Pons2013}.) The initially uniform internal temperature is 10$^{10}$\,K for all simulations, and we use vacuum outer boundary conditions. 

\subsection{Forces contributing to the NS core electric field}\label{section:allforces}

Here we wish to use the numerical simulations to investigate expulsion of magnetic flux from the core of neutron stars and to quantify the observable effects. First, we will discuss various mechanisms which are hypothesized to act on magnetic fluxtubes in the core and formulate a general, vectorial expression for the fluxtube drift velocity that we have incorporated into the numerical model. 

\cite{Muslimov1985} suggested that magnetic flux quanta should be subject to a buoyant force in regions where the NS core exhibits proton superconductivity. This buoyant effect is proposed to arise from the external material pressure gradient across each fluxoid surface (see \cite{Baym1975}). In the case of type-II superconducting protons, this buoyant force will also affect magnetic fluxtubes due to flux freezing between the magnetic field and proton fluid. The buoyant force per unit fluxtube length, $\vec{f}_b$, is directed radially outward with the form

\begin{equation}\label{equation:buoyancy1}
\vec{f}_b=-\frac{\mathcal{E}_f}{c_s^2}\vec{g}
\end{equation}

\noindent where $c_s$ is the local sound speed, $\vec{g}$ is the local gravitational acceleration, and
$\mathcal{E}_f=\left(\Phi_0/4\pi\lambda_p\right)^2\mathrm{ln}(\lambda_p/\xi_p)$ is the fluxtube energy per unit length. Here
$\Phi_0 = hc/2e \approx 2\times 10^{-7}$\,G\,cm$^{-2}$ is the quantum of magnetic flux
with $\xi_p$ and $\lambda_p$ being the coherence length and London penetration depth, respectively.
These characteristic length scales are approximated by $\xi_p\approx 6\times10^{-13}\rho_{p,13}^{-1/2}\Delta_p^{-1}$\,cm, and $\lambda_p=9\times10^{-12}\rho_{p,13}^{-1/2}$\,cm. Here, $\rho_{p,13}$ is the proton mass density in units of $10^{13}$\,g\,cm$^{-3}$ and $\Delta_p$ is the proton energy gap in MeV. Therefore, $\lambda_p / \xi_p \approx 15\Delta_p$, which for typical values of the gap fulfills the condition for type-II superconductivity $\lambda_p / \xi_p > 1/\sqrt{2}$. 
With these numerical estimates, an approximation to the fluxtube energy per unit length is $\mathcal{E}_f\approx 6\times 10^6\,\rho_{p,13}$\,erg\,cm$^{-1}$. The fluxtube energy per unit length is related to the lower critical field $H_{c1}$ by 

\begin{equation}
H_{c1}=\frac{4\pi\mathcal{E}_f}{\Phi_0}\approx 3.8\times 10^{14}\rho_{p,13}\,\mathrm{[G]}
\end{equation}

\noindent which is clearly a function only of density, and thus has only radial dependence in our spherical NS model. \cite{Muslimov1985} estimated that this buoyant force will draw the magnetic field out of the core proton fluid and into the sub-crustal region on time-scales of $\sim$50\,kyrs, where it may then be dissipated.

Scattering of the degenerate, relativistic electrons on the fluxoid field constitutes a viscous interaction in the NS core. This effect has been considered in previous studies with the drag coefficient being cast in a variety of ultimately equivalent forms. For our purposes we follow the notation of Alpar, Langer \& Sauls (1984) where we identify for simplicity the dimensionless drag coefficient $\mathcal{R}_p \approx 1.32 \times 10^{-2}\left(m_p / m_p^*\right)^{1/2}x_p^{1/6}\rho_{14}^{1/6}$ associated with scattering by a single fluxtube, where $x_p$ is the proton fraction, $m_p^*$ is the  proton effective mass, and $\rho_{14}$ is the mass density in units of $10^{14}$\,g\,cm$^{-3}$. This coefficient directly relates the force per unit fluxoid length to the relative velocity between the fluxtube and the electron fluid $\vec{u}_p-\vec{v}_e$. Specifically, the drag force due to scattering from a single fluxoid is 

\begin{equation}\label{equation:drag}
\vec{f}_D=-\kappa\rho_p\mathcal{R}_p\left(\vec{u}_p-\vec{v}_e\right)
\end{equation}

\noindent where $\kappa=h/2m$ is the quantum of circulation. It is currently unclear how the drag coefficient $\mathcal{R}_p$ behaves when derived in the nonlinear scattering limit, although \cite{Jones2006} suggested that $\mathcal{R}_p$ decreases with increasing fluxoid density, as in magnetars, where the electron mean free path is larger than inter-fluxoid spacing.

If fluxoids in the NS core deviate from a perfectly straight configuration, then a self-tension force will act to release the field curvature. \cite{Alpar1984} suggested that this mechanism is non-negligible even in the case of weak $\sim$10$^{12}$\,G fields. Harvey, Ruderman \& Shaham (1986) included this tension force in their analysis of core field decay, as did \cite{Konenkov2000}, although only at the crust-core interface. \cite{Konenkov2000} reported that this inclusion delayed the expulsion of magnetic flux significantly, as our simulation results also predict. The self-tension force is given by

\begin{equation}\label{equation:tension}
\vec{f}_T = \mathcal{E}_f\left(\hat{b}\cdot\vec{\nabla}\right)\hat{b}
\end{equation}

\noindent where $\hat{b}$ is the unit vector in the local magnetic field direction.  

\cite{Jones1987} suggested that the Magnus force acting in the superfluid, superconducting NS core can significantly delay the expulsion of magnetic flux, contradicting earlier, faster, expulsion rate estimates. In addition, the Magnus force can act to twist the core magnetic field due to a strong azimuthal component. Indeed, Andersson, Sidery \& Comer (2006) and \cite{Glampedakis2011} arrived at similar conclusions by including this effect in their calculations. Following these earlier works, the Magnus force per unit length acting on single fluxtube is given by

\begin{equation}\label{equation:magnus1}
\vec{f}_M=-{\rho_p \kappa }\left(\vec{u}_p - \vec{v}_e\right)\times \hat{b}
\end{equation}

\noindent where the local background velocity is assumed to be the electron velocity $\vec{v}_e$. 


\subsection{The induction equation in the NS core}\label{section:fullforcebalance}

We now combine all of the acting forces discussed in this section and derive a vectorial expression for the effective drift velocity of magnetic fluxoids in the simulated NS core. Adding equations (\ref{equation:buoyancy1}), (\ref{equation:drag}), (\ref{equation:tension}), and (\ref{equation:magnus1}) the total force balance equation is 

\begin{equation}
\vec{f}_*=\vec{f}_b+\vec{f}_D+\vec{f}_T+\vec{f}_M= 0
\end{equation} 

\noindent which gives the following vector expression relating the components of the drift velocity, the magnetic field components, and fundamental constants:

\begin{equation}\label{equation:vectoru1}
\mathcal{R}_p\vec{u}+\left(\vec{u}\times\hat{b}\right)=\frac{\mathcal{E}_f}{\kappa\rho_p}\left[-\frac{\vec{g}}{c_s^2}+\left(\hat{b}\cdot\vec{\nabla}\right)\hat{b}\right]
\end{equation}

\noindent with $\vec{u}\equiv\vec{u}_p-\vec{v}_e$. 

The explicit solution for $\vec{u}$ in terms of the RHS of equation (\ref{equation:vectoru1}), hereafter denoted by $\vec{V}$, 
may be written as in \cite{Hall1956}

\begin{equation}\label{equation:vectoru2}
\vec{u}=\frac{1}{1+\mathcal{R}_p^2}\left[\mathcal{R}_p\vec{V}+\hat{b}\times\vec{V}+\frac{\left(\hat{b}\cdot\vec{V}\right)\hat{b}}{\mathcal{R}_p}\right]
\end{equation}

This solution permits drift components parallel and perpendicular to the local magnetic field.  
We limit our formulation to consider only the perpendicular part, because only perpendicular components will displace the core field. Removing the parallel drift terms, the mesoscopic fluxtube velocity is

\begin{equation}\label{equation:vectoru3}
\vec{u}_{\bot}=\frac{1}{1+\mathcal{R}_P^2}\left[\mathcal{R}_p\vec{V}_{\bot}+\hat{b}\times\vec{V}\right]
\end{equation}

\noindent and where the source term $V_{\bot}$ may be more instructively written as

\begin{equation}\label{equation:vectoru4}
V_{\bot}=\frac{\mathcal{E}_f}{\kappa\rho_p}\left[-\hat{b}\times\left(\left(\vec{\nabla}+c_s^{-2}\vec{g}\right)\times\hat{b}\right) \right]
\end{equation}

\noindent which highlights how the release of field tension is modified by the presence of buoyant forcing. The result is that the tension force is weakened, yielding an effective radius of curvature that is larger in regions were the buoyant forcing is strongest, i.e. in the equatorial region. The effect is thus a reduction in the local restoring force. In the same spirit as \cite{Glampedakis2011a} and \cite{Graber2015}, the core electric field from equation (\ref{equation:coreEfield1}) may then be written as

\begin{equation}\label{equation:coreEfield2}
\vec{E}=-\frac{1}{c}\vec{v}_e \times\vec{B}-\frac{1}{e}\vec{\nabla}\left(\mu_e+m_e\Phi\right)
+\frac{B}{c}\mathcal{R}_p\vec{u}_{\bot}.
\end{equation}

\noindent The second term on the RHS, containing chemical potential and gravitational potential gradients, vanishes exactly upon insertion of equation (\ref{equation:coreEfield2}) into equation (\ref{equation:MTinduction}). Owing to the extremely weak London field, the electrons and protons are co-moving in the superconducting core \citep{Glampedakis2011a}. Thus we can further neglect the first RHS term in the non-rotating case, and the final induction equation in our simulated NS core is

\begin{equation}\label{equation:coreinduction}
  \partial_t\vec{B}=-\vec{\nabla}\times\left(B\mathcal{R}_p\vec{u}_{\bot}\right).
\end{equation}

To summarize, in the NS crust we advance the magnetic field by computing the Hall-Ohmic electric field (\ref{equation:crustinduction}) and then update the magnetic field in time with equation (\ref{equation:MTinduction}); the components of the fields and currents are computed on a staggered grid, using the two-step scheme described by \cite{Vigano2012}. In the NS core, the magnetic field is evolved with equation (\ref{equation:coreinduction}), where the local fluxtube drift $\vec{u}_{\bot}$ is computed on the numerical grid with equation (\ref{equation:vectoru3}), which requires the calculation of local gravitational acceleration and field tension at each numerical cell. We advance both the crustal and core temperatures by stepping equation (\ref{equation:MTtemperature}) forward in time.

\subsection{Energy Considerations}\label{section:EnergyConsiderations}

In normal matter, the expression for local energy balance between fields and currents is

\begin{equation}\label{eq:ejconservation}
\partial_t\left(\frac{B^2}{8\pi}\right)=-\vec{E}\cdot\vec{J}-\vec{\nabla}\cdot \vec{S}
\end{equation}

\noindent where on the LHS we identify the time rate of change of the local magnetic energy density, $\partial_t\,\epsilon_B$ hereafter,
and $\vec{S}=c\vec{E}\times\vec{B}/4\pi$ is the Poynting vector. In the crust we have two contributions to the electric field, so following from equation (\ref{equation:MTinduction}), the instantaneous local energy balance is given by

\begin{equation}\label{equation:pwconservation}
\partial_t\,\epsilon_B = -Q_j-\vec{\nabla}\cdot \vec{S}
\end{equation}

\noindent where $Q_j=4\pi \eta J^2 / c$ is the Joule heating rate per unit volume. Integrating over the crust volume 
we obtain the required balance between energy rates in the NS crust, with the crust-core interface designated by $R_c$: 

\begin{equation}\label{eq:totalconservation}
d_t\,E_B=-\int_{R_c}^{R_*}\!\!\!\!\mathrm{d}V Q_j-\int_{\partial R_*} \!\!\!\!\mathrm{d}\vec{A}_r\cdot\vec{S}
\end{equation}

\noindent In practice, the Poynting losses at the outer numerical domain (the stellar surface $R_{*}$) are neglected, because we impose vacuum boundary conditions ($\vec{E}=0$ for a non-rotating star). In the absence of self-consistent magnetospheric feedback, this approach is standard practice. Joule dissipation dominates the magnetic energy losses, which is enhanced by the Hall cascade. Integrating equation (\ref{eq:totalconservation}) in time we obtain the time-integrated energy conservation requirement at arbitrary time $t$:

\begin{equation}\label{eq:timeconservation}
E_B(t)=E_B(0)-\int\mathrm{d}t\left(Q_{j,\mathrm{tot}}+\mathcal{F}_{S,\mathrm{tot}}\right)
\end{equation}

Now we investigate how the time-dependent energy balance is affected in the core by the fluxoid drift velocity prescribed in equation (\ref{equation:vectoru3}). Following \cite{Graber2015}, the magnetic energy density in the superconducting core is $\epsilon_{sc}=\vec{H}_{c1}\cdot\vec{B}/2\pi$, where $\vec{H}_{c1}=H_{c1}\hat{b}$ is always parallel to the local macroscopic magnetic field. As already pointed out in that paper, the time rate of change in the local magnetic energy density from equation (\ref{equation:MTinduction}) couples to the evolution of matter in the NS core, via

\begin{equation}\label{eq:sclocal1}
\partial_t\,\epsilon_{sc}=\frac{1}{2\pi}\vec{H}_{c1}\cdot\partial_t\vec{B} + \frac{1}{2\pi}\vec{B}\cdot\partial_t\vec{H}_{c1}.
\end{equation}

\noindent Neglecting the evolution of matter in the second term on the RHS, we may write local energy conservation as a combination of surface and bulk terms, as

\begin{equation}\label{equation:scfull}
\partial_t\,\epsilon_{sc}=\vec{\nabla}\cdot\vec{\Sigma}-\frac{\epsilon_{sc}\mathcal{R}_p\mathcal{E}_f}{\rho_p\kappa\left(1+\mathcal{R}_p^2\right)}\left(\mathcal{J}_{\bot}^2+\frac{\vec{\nabla}\rho_p}{\rho_p}\cdot\vec{\mathcal{S}}_B-\mathcal{W}_B\right)
\end{equation}

\noindent where we have used the shorthand $\vec{\mathcal{J}}=\vec{\nabla}\times\hat{b}$ in the dissipative Joule-like term presented by \cite{Graber2015}. The second bulk term on the RHS accounts for coupling to the density stratification $\vec{\nabla}\rho_p$, being purely radial and unique to the equation of state. Physically, this term admits drift waves in the superconducting medium, with the imposed density gradient supplying the energy. As suggested in equation (\ref{equation:vectoru4}), it is now convenient to take the shorthand $\vec{\mathcal{J}}_g=\vec{g}_{\bot}\times\hat{b}/c_s^2$. Then, an interesting parity is found where gravity plays a role not unlike a differential operator:  

\begin{equation}
\vec{\mathcal{S}}_B=\left(\mathcal{R}_p\vec{\mathcal{J}}-\vec{\mathcal{J}}\times\hat{b}\right)+\left(\mathcal{R}_p\vec{\mathcal{J}}_g+\vec{\mathcal{J}}_g\times\hat{b}\right)
\end{equation}

\noindent The final RHS bulk term in equation (\ref{equation:scfull}) arises solely from the presence of buoyancy, and represents the local work done against gravity by the drifting fluxtubes: 

\begin{equation}
  \mathcal{W}_B=-\vec{\mathcal{J}}\cdot\left(\mathcal{R}_p\vec{\mathcal{J}}_g+\vec{\mathcal{J}}_g\times\vec{b}\right).
\end{equation}

Despite accounting explicitly for all terms expected to arise in the energy balance, there are legitimate computational challenges. It is far beyond the scope of this work to realistically model e.g. the thermodynamics associated with the superconducting-normal matter phase transition at the crust-core interface, or detailed superconducting microphysics that our model is simply not designed for. However because our primary goal is to gain an understanding for the expulsion of magnetic energy--from the NS core into the crust--we will limit our attention to estimates of such rates, which will be discussed in parallel with the simulation results in \S\ref{section:simulationresults}. Thus we will defer the computation of the myriad bulk contributions to the NS core energy balance at the moment and instead focus on the surface term in equation (\ref{equation:scfull}). The local energy flux density at the crust-core interface is

\begin{equation}\label{equation:surfaceterm}
\vec{\Sigma}=\epsilon_{sc}\mathcal{R}_p \vec{u}_{\bot}\times\hat{b}
\end{equation}

\noindent where $\vec{u}_{\bot}$ is shown in equation (\ref{equation:vectoru3}). We monitor this diagnostic in volume-integrated form in conjunction with equation (\ref{equation:scfull}):

\begin{equation}\label{equation:scboundarycons}
\frac{d E_{sc}}{dt} = \int_{\partial R_c}\mathrm{d}\vec{S}\cdot\vec{\Sigma}
\end{equation}

\begin{figure}
\centering
\includegraphics[width=8.5cm,height=7.23cm]{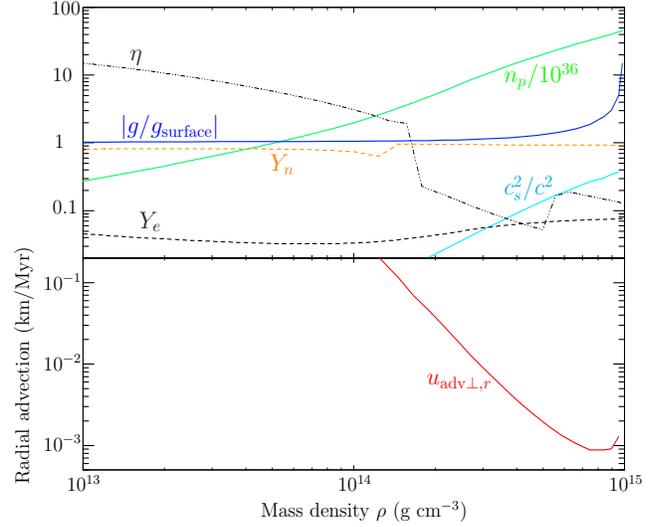}
\caption{{\bf Top:} Proton number density $n_p$, gravitational acceleration $|g|$, sound speed $c_s^2$, neutron fraction $Y_n$, electron fraction $Y_e$, $t=0$ electrical resistivity $\eta$ in units of km$^2$\,Myr$^{-1}$. {\bf Bottom:} Estimate of radial fluxtube drift speed.}
\label{fig:eosparams}
\end{figure}

\section{Analytic Estimates}\label{section:analyticestimates}

\begin{figure*}
\centering
\includegraphics[width=\textwidth]{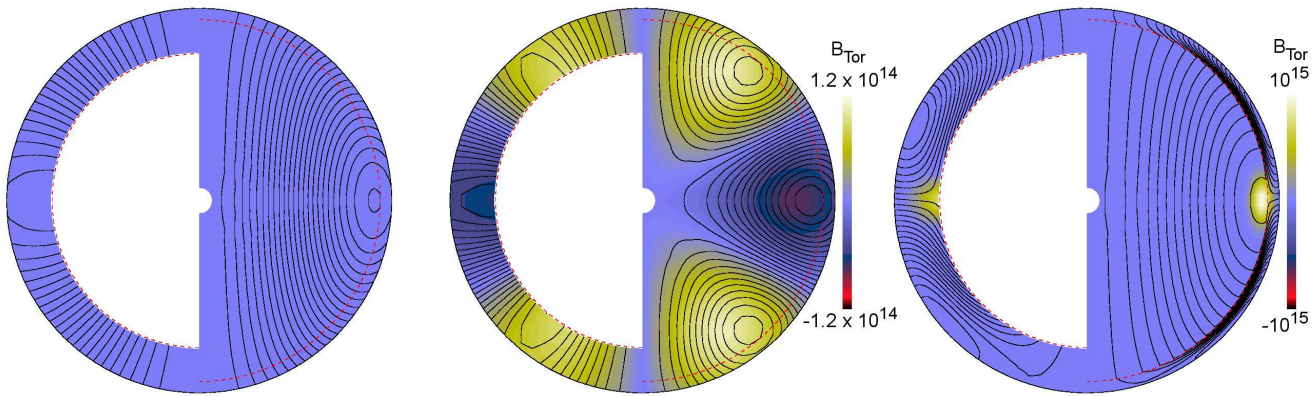}
\caption{Initial magnetic field geometries for our three simulations. {\bf Left:} core-extended purely poloidal dipole. {\bf Middle:} core-extended octopole. {\bf Right:} large-scale dipole with crust-confined dipolar component. Left half of each plot shows magnified crust for visualization aid. Red dotted lines indicate crust-core interface.}
\label{fig:Binit}
\end{figure*}

Here we discuss our expectations from analytic estimates regarding the expulsion rate and the net effects on neutron star evolution, and how we may connect these predictions with observational data. Let us consider the behavior associated with the drift components given in equation (\ref{equation:vectoru3}). First, we imagine for simplicity that the magnetic field which threads the core is perfectly aligned with the magnetic axis, i.e., the toroidal field component $b_{\phi}$ is initially zero. Let us also ignore the field self-tension for a moment; this term delays the buoyant drift, so omitting this tension here will over-estimate the flux expulsion rate. Typically, $\mathcal{R}_p\approx 10^{-2}$ (\cite{Alpar1984, Glampedakis2011}), so near the magnetic equator the drifts are $u_r \approx \mathcal{E}_f\mathcal{R}_p g/\kappa\rho_p c_s^2$, $u_{\theta}=0$, $u_{\phi}\approx-u_r/\mathcal{R}_p$ because $b_{\theta}\approx 1$. Thus, in such a poloidal configuration the magnetic field will twist up azimuthally on a time-scale of $1/\mathcal{R}_p$ faster than the radial displacement time-scale, which is a direct consequence of including the Magnus force in our formulation. That is, the Magnus force redistributes radial momentum into angular flows, such that the radial buoyant drift is converted into polar and azimuthal drifts, reducing radial expulsion rates in the linear regime. Inserting nominal values from a particular NS model (top of figure \ref{fig:eosparams}), we obtain the approximate radial fluxtube drift speed shown in the bottom of figure \ref{fig:eosparams}. If we consider instead that the initial magnetic field is primarily toroidal in the region of interest, then by construction a strong $u_{\phi\bot}$ will not initially be present. As in the purely poloidal case, $u_r \approx \mathcal{E}_f\mathcal{R}_p g/\kappa\rho_pc_s^2$, but the dominant component of fluxtube motion is now  equatorward with $u_{\theta}\approx u_r/\mathcal{R}_p$. Thus, regardless of the initial conditions in the NS core region, the system should relax to (or begin in) a state where a strong toroidal component is present, which based on the weak Joule dissipation should persist for a majority of the observable lifetime.

Directing attention to the drift term from the induction equation (\ref{equation:MTinduction}), obtaining the characteristic fluxtube drift time-scale $t_{\mathrm{dr}}\sim L_{\mathrm{dr}}/\mathcal{R}_p u_{\bot}$ is straightforward. From above, we may make the approximation $\mathcal{E}_f \approx \kappa^2\rho_p/4\pi$, and thus $u_r\approx\kappa\mathcal{R}_p/4\pi L_{\mathrm{dr}}$ where $L_{\mathrm{dr}}\sim c_s^2/g$ is the pressure length scale. The drift time-scale $t_{\mathrm{dr}}$ follows, and a comparison with the Ohmic dissipation time-scale in the core (Haensel, Urpin \& Yakovlev 1990; Goldreich \& Reisenegger 1992) gives


\begin{equation}\label{equation:advOhmTimescale}
\frac{t_{\rm{dr}}}{t_{\rm{Ohm}}}\sim{0.1}\frac{L_{\mathrm{dr}}^2}{L_{\mathrm{Ohm}}^2}\frac{T_{10}^2}{\mathcal{R}_p^2}\left(\frac{\rho_{\rm{nuc}}}{\rho}\right)^3
\end{equation}

\noindent where $L_{\mathrm{Ohm}}$ is the characteristic length scale of field variations in the NS core, $T_{10}$ is the temperature in units of 10$^{10}$\,K, and $\rho_{\rm{nuc}}$=2.8$\times$10$^{14}$\,g\,cm$^{-3}$. Because there is spatial variation in the imposed velocity field, and because the field in the outer core will be removed first (but possibly resupplied from the inner core), we first consider the time-scale ratio in the outermost core region. Only the radial component of $\vec{u}_{\bot}$ contributes to removal of magnetic flux from the core, and $L_{\mathrm{dr}}\sim$1\,km is an obvious length-scale. From figure \ref{fig:eosparams}, the radial velocity $u_{\bot}$ in such a region is $\sim$0.1\,km Myr$^{-1}$. Using $\rho\approx\rho_{\rm{nuc}}$, $\mathcal{R}_p\sim 10^{-2}$, and $L_{\mathrm{Ohm}}\sim$1\,km, we obtain $t_{\mathrm{dr}}\,\sim\,10^3\,T_{10}^2\,t_{\mathrm{Ohm}}$. The average core temperature reaches $10^9$\,K after $\sim$100\,yrs, and thus when the star is young and very hot ($>$\,10$^9$\,K) the fluxtube drift is slow compared to the Ohmic dissipation. After the NS cools, the prescribed core forcing becomes comparable to, and later, faster than Ohmic dissipation in the core for the full NS lifetime.

It is useful to compare the drift time-scale in the outer core with the Hall time-scale in the crust $t_{\mathrm{Hall}}\sim4\pi n_e e L_{\mathrm{Hall}}^2 / cB$. This will give an approximate metric for how quickly the magnetic field is redistributed in the crust versus the supply rate due to expulsion from the core. Using $L_{\mathrm{Hall}}\sim$1\,km (crust thickness), one obtains

\begin{equation}
\frac{t_{\mathrm{dr}}}{t_{\mathrm{Hall}}}\sim40\,B_{12}
\end{equation}

\noindent where $B_{12}$ is the magnetic field in units of 10$^{12}$\,G. Therefore the drift time-scale for the core fluxtube array will always be considerably slower than the crustal Hall cascade. As the Hall effect drives the field down to smaller spatial scales, where it is more efficiently dissipated through Joule heating, the transport of new field energy from the core into the crust should not be sufficient to replenish the crust with new magnetic field. We therefore expect that any extra field component confined to the crust will be dissipated through the combined action of the Hall and resistive terms before significant field energy is displaced into the crust from the core. We verify this expectation in \S\ref{section:simulationresults} by testing a variety of initial conditions.

\section{Simulation Results}\label{section:simulationresults}

We assess the impact of the macroscopic fluxtube drift by carrying out three sets of simulations which all use the drift prescription in equation (\ref{equation:vectoru3}).\footnote{We have performed complementary simulations which consider each of the forces from \S\ref{section:allforces} independently and in various combinations, but the results do not differ significantly from those presented here.} Each set is characterized by the prescription of the initial magnetic field configuration. The simplest ideal case is discussed in \S\ref{section:coreext}, where the initially poloidal magnetic field is taken to be dipolar with continuous crust-core matching. A second case presented in \S\ref{section:initff} takes initial conditions with a purely octopolar field, which has comparable poloidal and toroidal strengths. Here the primary defining feature is the presence of higher-order multipole structure in the NS interior. In \S\ref{section:inithybmix}, a crust-confined poloidal dipole is superimposed on the core-extended configuration, with a strong toroidal field in the equatorial region. 

First in each section, data illustrating the overall energy budget are shown, displaying the volume-integrated instantaneous rate of magnetic energy decay ($|d E_b/dt|$), the decay rate in the superconducting core ($+d E_{sc}/dt$), the Joule rate $Q_j$, the magnetic energy growth rate in the core ($-d E_{sc}/dt$), and flux of magnetic energy across crust-core interface $Q_{\mathrm{flux}}$. $+Q_{\mathrm{flux}}$ indicates magnetic energy transport from the core to the crust, and $-Q_{\mathrm{flux}}$ indicates transport in the opposite direction. Second, we present the cumulative total energy balance in each reservoir. Snapshots of the fluxtube drift velocity, magnetic field, and current intensity are also shown at characteristic times during the evolution. There, poloidal magnetic fields are shown superimposed over the radial and azimuthal fluxtube drift components. A discussion of the simulated magnetic field evolution in the core is provided in \S\ref{section:comments}.

\subsection{Core-extended dipole}\label{section:coreext}

\begin{figure}
\centering
  \includegraphics[width=8cm,height=7.5cm]{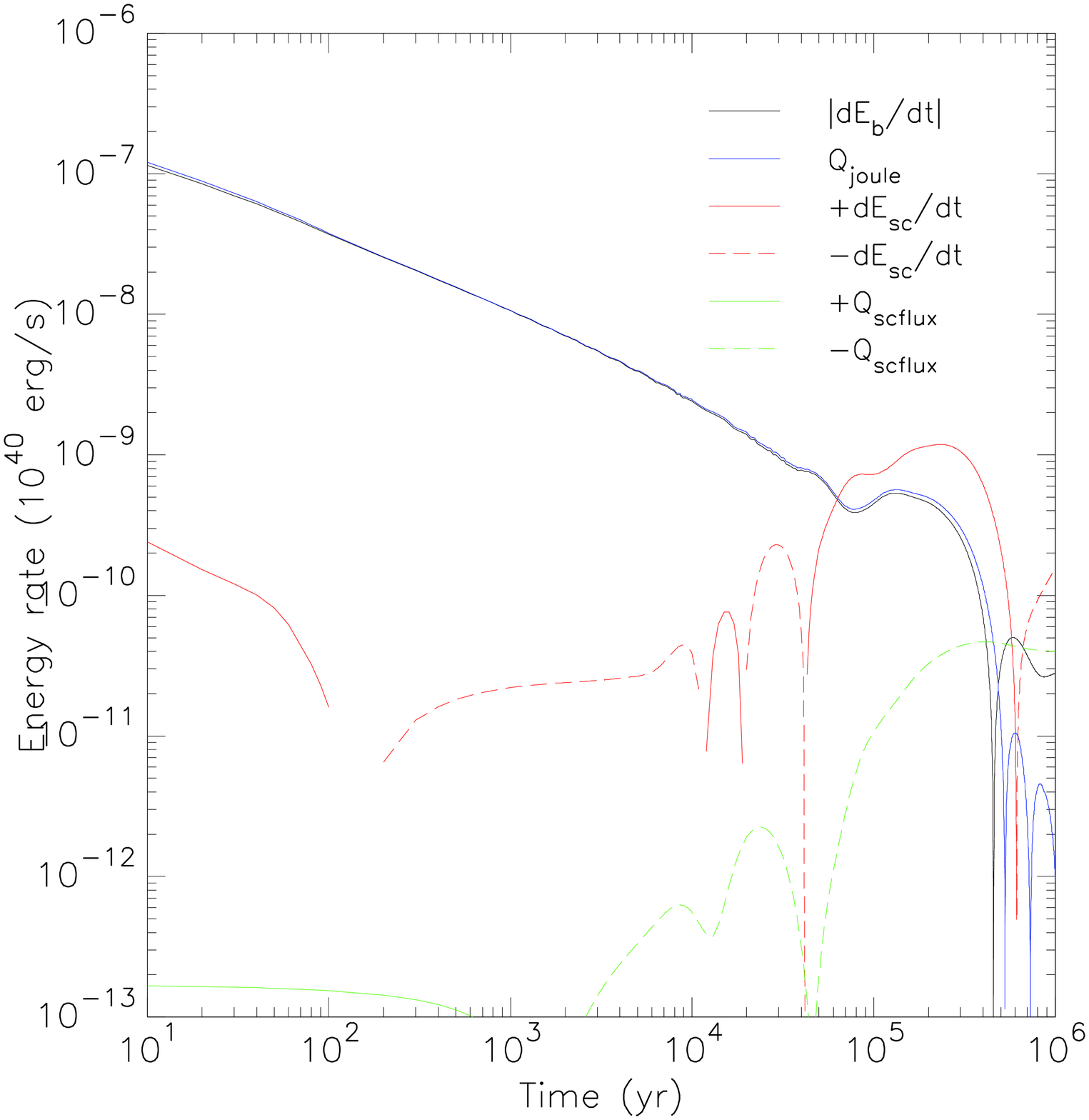}
  \caption{Energy budget in core-extended configuration. Total magnetic energy decay (black), superconducting magnetic energy decay/growth in core (solid/dashed red), losses from Joule dissipation (blue), radially outward/inward flux of magnetic energy at crust-core interface (solid/dashed green).}
  \label{fig:coreextbcons}
\end{figure}

Figure \ref{fig:Binit} (left) shows the initial magnetic field configuration. The initial magnetic field is a purely poloidal dipolar configuration with strength 10$^{14}$\,G at the pole, and threads the core and crust uniformly. This is commonly referred to as a core-extended configuration. A discussion of the energetics follows. 

\begin{figure}
\centering
\includegraphics[width=8cm,height=7.0cm]{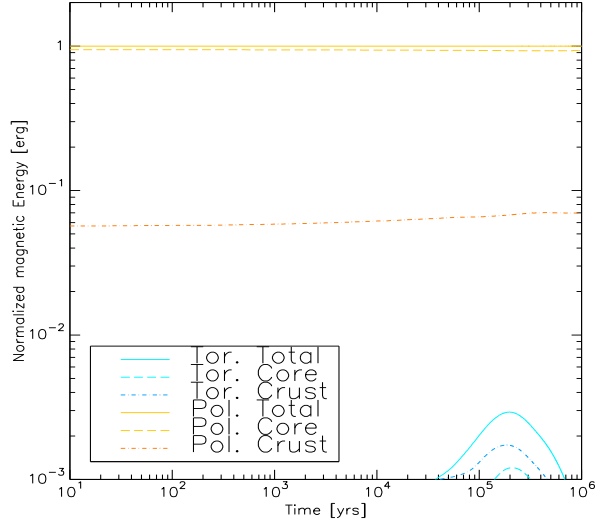}
\caption{Magnetic energy normalized to instantaneous total magnetic energy $E_b(t)$, in core-extended configuration.}
\label{fig:coreextBenergy}
\end{figure}

\begin{figure*}
  \includegraphics[width=\textwidth]{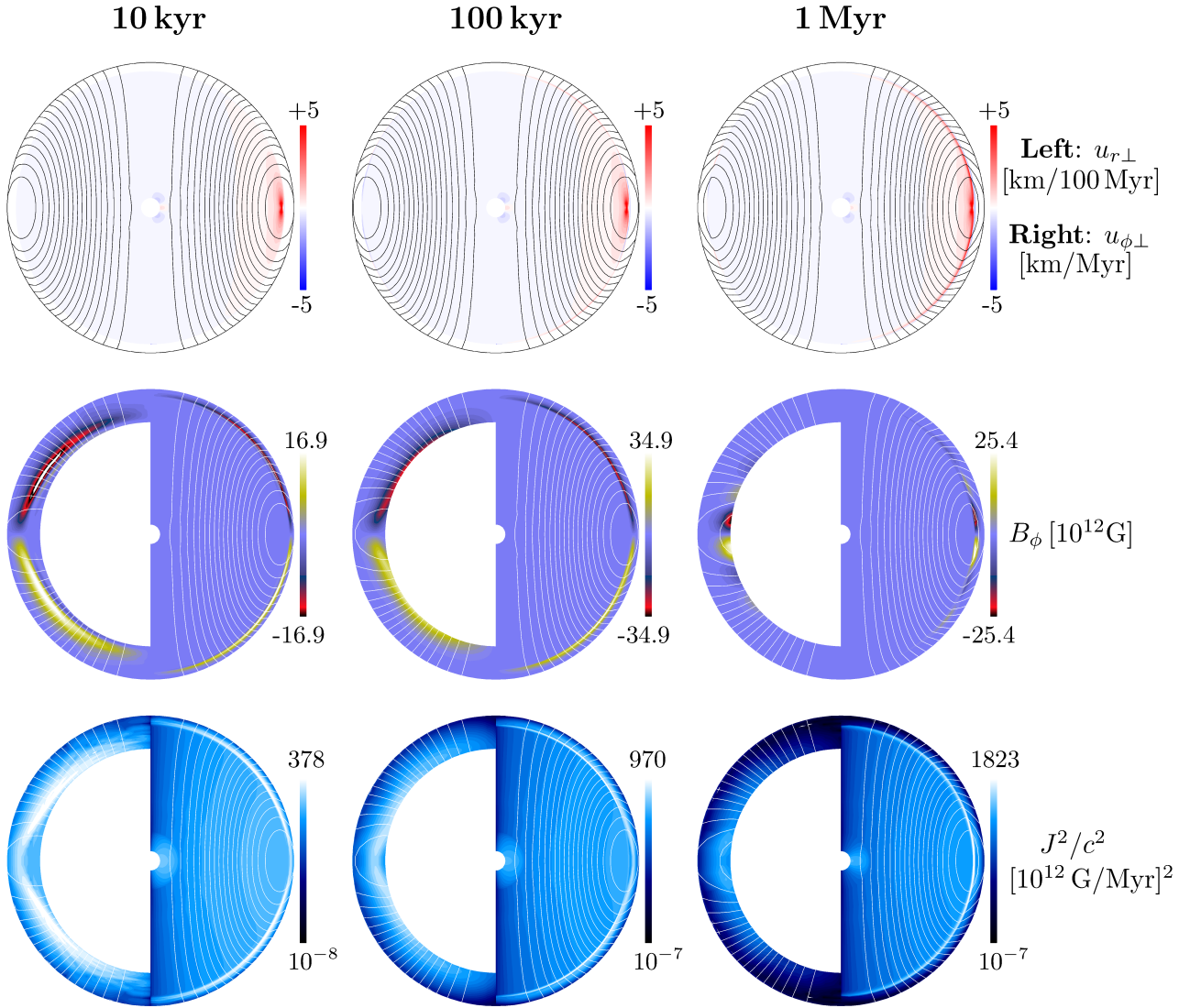}
  \caption{Snapshots of magnetic field diagnostics, core-extended case. Poloidal magnetic field lines are superimposed on $u_{\bot}$ (top), $B_{\phi}$ (middle) and current intensity $J^2$ (bottom). {\bf Top row:} left half shows $u_{r\bot}$, right half shows $u_{\phi\bot}$. {\bf Middle, bottom rows:} Crust thickness exaggerated for visualization purposes in left half of each panel.}
\label{fig:cextbjmaps}
\end{figure*}

From figure \ref{fig:coreextbcons}, Joule dissipation throughout the star is primarily responsible for the decay of magnetic energy, with the dissipation in the NS core being insignificant until at least 50\,kyrs given these idealized initial conditions. One immediately obvious feature is the oscillatory behavior in the energy of the superconducting core; this arises from a combination of factors. See \S\ref{section:comments} for a generalized discussion. For the first 100\,kyrs of the evolution, there is a weak submergence of magnetic energy from the crust into the core, which comprises a non-negligible fraction of the total energy budget after 100\,kyrs. In addition, an increase in the superconducting magnetic energy decay rate is noted at this time. Beyond a few hundred kiloyears, the global magnetic energy decay rate decreases to $\dot{E}_b < 10^{30}$\,erg\,s$^{-1}$, while the transport of magnetic energy from the crust to the core remains roughly constant for the final 500\,kyrs. Overall, the total stellar magnetic field energy decays very weakly after this time, indicating a globally static magnetic field topology.

Such weak decay is further evidenced by considering each reservoir of volume-integrated magnetic energy, specifically the instantaneous energy balance of the poloidal and toroidal components of the core and crustal fields. The evolution of each volume-integrated component is shown in figure \ref{fig:coreextBenergy}, where the energies are normalized to the time-dependent total magnetic energy. The benefit of this, instead of normalizing to the initial energy, is to systematically remove the dissipated field energy and thus focus more easily on the migration of field energy across the crust-core interface. Immediately one recognizes that the induction of toroidal field components in the core and crust occurs on a time-scale of 100\,kyrs. In the NS crust, the Hall cascade is responsible, whereas in the core the toroidal component arises due to field lines experiencing differential drift in the azimuthal direction. Both core and crust toroidal components reach similar values, comprising 0.1 per cent of the magnetic energy at $\sim$100\,kyrs. 

Structure of the fluxtube velocity (top), magnetic fields (middle), are current intensity (bottom) are shown in figure \ref{fig:cextbjmaps} at key times. Induction of a toroidal field component accompanies the outer crust-confined currents within the first few kiloyears, as expected, via the Hall-driven evolution of poloidal field. The system eventually reaches a quasi-equilibrium state as the primary current sheet submerges through the crust and straddles the crust-core interface at later times. By 100\,kyrs the interface is the most important location for the ensuing evolution, as Joule heating in this region will control the global dissipation rates. Within the first 100\,kyrs no significant deviation of the poloidal field configuration away from the initial conditions is observed. As predicted, the fluxtube motion is fastest in the outermost NS core, dominated by a strong azimuthal component (top row, right half of each plot) that is two to three orders of magnitude faster than the radial fluxtube drift (top row, left half of each plot). Both the azimuthal and radial drift speeds increase with time, owing to the enhanced magnetic field tension near the interface as the crustal field decays and thus intensifies field line tension.

\subsection{Core-extended octopole}\label{section:initff}
\begin{figure}
\centering
  \includegraphics[width=8cm,height=7.5cm]{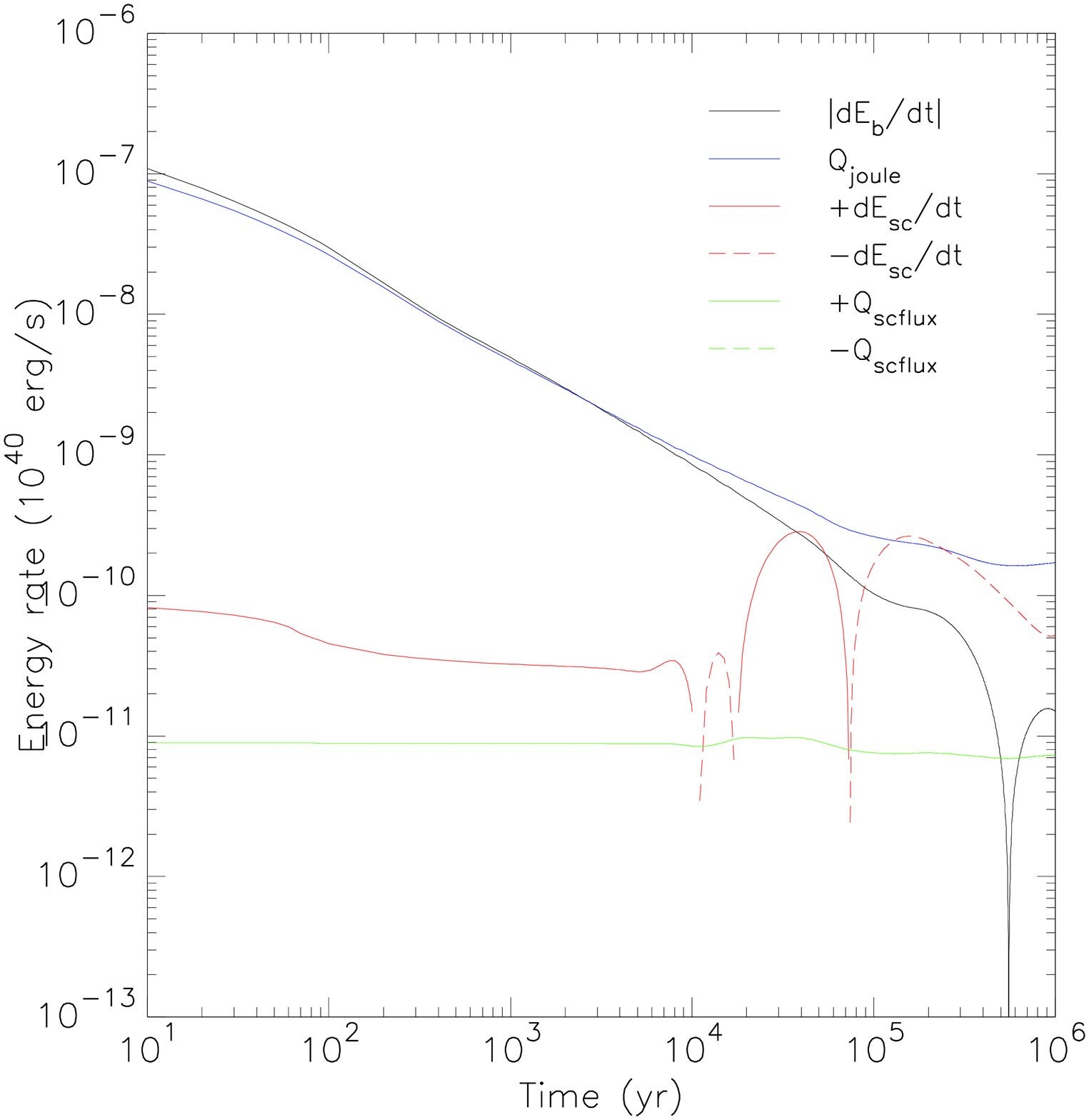}
  \caption{Energy budget in octopolar configuration. Total magnetic energy decay (black), superconducting magnetic energy decay/growth in core (solid/dashed red), losses from Joule dissipation (blue), radially outward/inward flux of magnetic energy at crust-core interface (solid/dashed green).}
  \label{fig:ffbcons}
\end{figure}

It is also useful to consider initial conditions in the core which exhibit higher-order structure compared to the simplest dipolar scenario presented \S\ref{section:coreext}. Figure \ref{fig:Binit} (middle) shows the magnetic field initial conditions for the case of a octopole. It is clear that this magnetic topology exhibits considerable field-line tension in the NS core. Initially, the peak poloidal field magnitude is $10^{14}$\,G, and the peak toroidal field is $1.2\times10^{14}$\,G. Contributions to the energy balance are examined in figure \ref{fig:ffbcons} for the instantaneous, volume-integrated power. An immediate comparison with the core-extended case (figure \ref{fig:coreextbcons}) reveals that given these initial conditions, the core magnetic field is expelled into the crust during the first megayear, albeit weakly. The global magnetic energy decay is again secular, as in the purely dipolar case, and both obey the same power law for the first 100\,kyrs of evolution. Fluctuations in the superconducting field energy decay are again observed, as with the dipolar case (see \S\ref{section:comments} for a discussion).

\begin{figure}
\centering
\includegraphics[width=7cm,height=6cm]{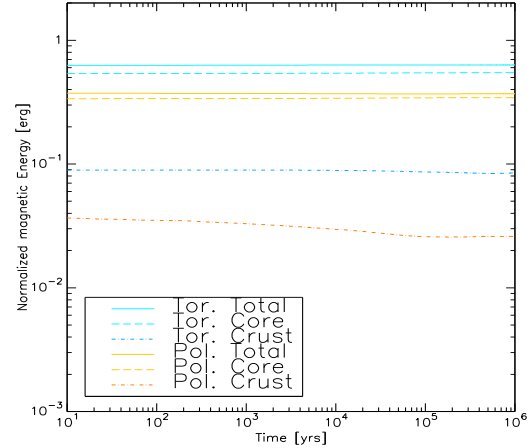}
\caption{Magnetic energy normalized to instantaneous total magnetic energy $E_b(t)$, in octopole configuration.}
\label{fig:ffBenergy}
\end{figure}

When imposing these initial conditions, the crustal magnetic field evolves such that a current sheet is induced in the outermost crust, due to the discontinuity in $B_{\phi}$ at the NS surface (the vacuum magnetosphere has no toroidal component). It is expected that explicitly imposing a global fluxtube drift would accelerate a deviation from the initial configuration in the NS core, but surprisingly, the partition of energy between toroidal and poloidal components is time-independent in the core, illustrated in figure \ref{fig:ffBenergy}. Some crustal field energy is dissipated by 100\,kyrs, but again the slow expulsion cannot redistribute magnetic energy confined to the core on sub-megayear time-scales. 

\begin{figure*}
  \includegraphics[width=\textwidth]{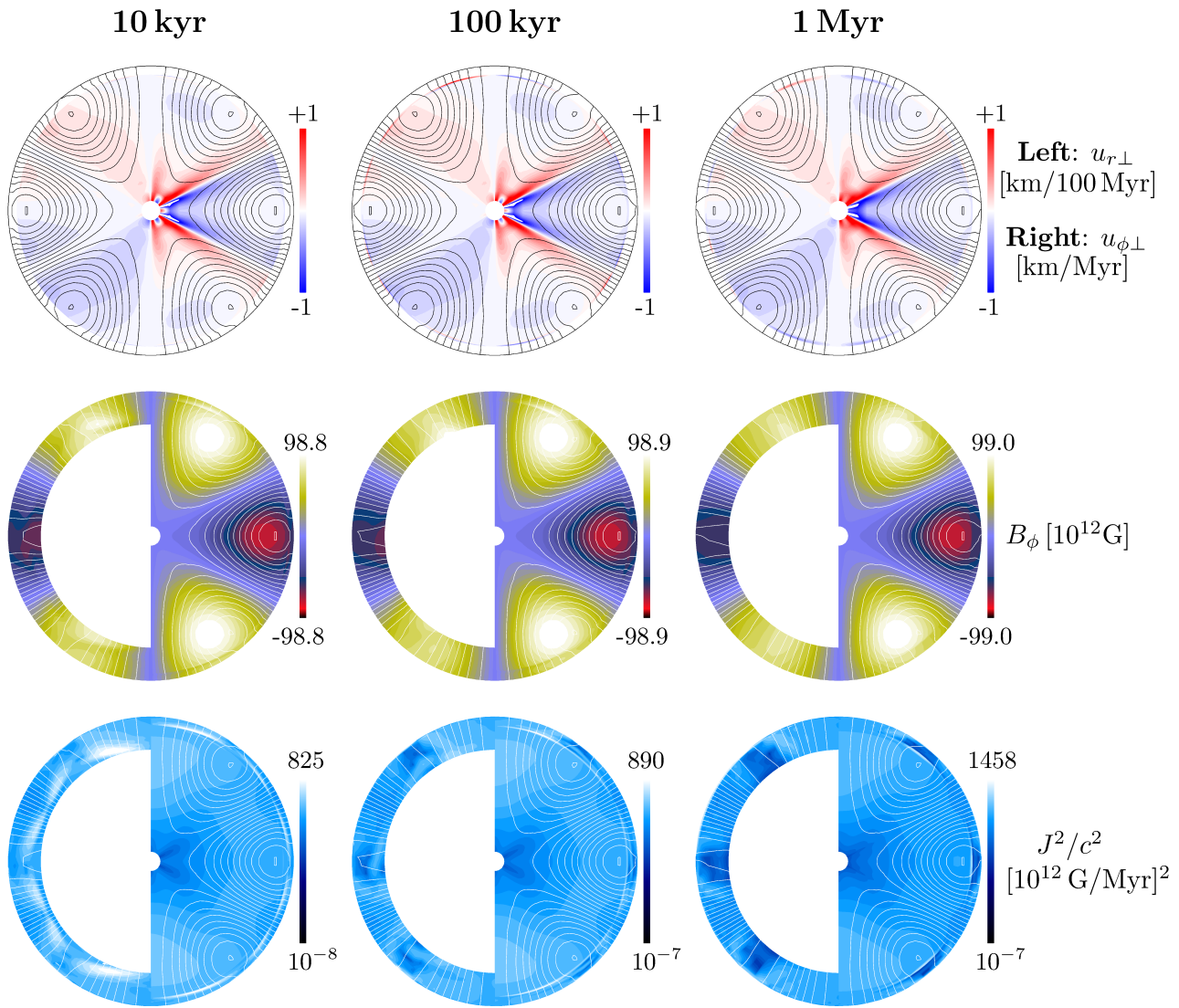}
    \caption{Snapshots of magnetic field diagnostics, octopolar case. Poloidal magnetic field lines are superimposed on $u_{\bot}$ (top), $B_{\phi}$ (middle) and current intensity $J^2$ (bottom). {\bf Top row:} left half shows $u_{r\bot}$, right half shows $u_{\phi\bot}$. {\bf Middle, bottom rows:} Crust thickness exaggerated for visualization purposes in left half of each panel.}
\label{fig:ffbjmaps}
\end{figure*}

Figure \ref{fig:ffbjmaps} shows evolution and decay of the global magnetic field and currents, along with fluxtube drift maps. As usual, the field configuration in the crust control dynamics up to 10\,kyrs, where additional toroidal components are induced and where the currents are localized. Kinking of the poloidal field lines near the interface is observed within the first 100\,kyrs, supported by substantial currents at the interface region, but afterward a primarily radial field persists in the NS crust. In the core-extended case the strongest azimuthal drift was isolated to the magnetic equator at the crust-core interface, but here the nature of the global drifts is more structured. The drift patterns follow the field topology, with the peak azimuthal drifts corresponding with regions of maximum magnetic tension and zero toroidal field (by design, see equation (\ref{equation:vectoru3})). Since the initial toroidal field is inherently twisted, an off-equatorial azimuthal shear flow is present in the core, which persists throughout the entire simulation. Thus, high-order multipolar structure, or even turbulent internal structure, will produce topologies which are potentially sensitive to tearing instabilities and magnetic reconnection in NS cores. The hemispheric discrimination in $u_{r\bot}$ is an unexpected feature here. This north-south asymmetry in the radial fluxtube drift arises from the self-consistent toroidal component. The lack of a radial drift near the equator is a direct consequence of this asymmetry. Thus in this configuration any field energy expelled from the core would boost the crustal field energy is these off-equatorial regions. However, we also note that the color scale here is considerably smaller than in the core-extended dipolar case, evidence that the expulsion would be even slower in this higher multipole, higher tension model. 

\subsection{Crust-confined dipole}\label{section:inithybmix}

Figure \ref{fig:Binit} (right) shows the magnetic field initial conditions for a large-scale dipolar magnetic configuration with a strong dipolar component confined to the NS crust. The toroidal field is confined to the last closed field line region saddling the crust-core interface, with magnitude $10^{15}$\,G. The initial poloidal field strength at the pole is $10^{14}$\,G. The crust-confined dipole adds a significant amount of magnetic energy in the crust; such configurations have been extensively used to unify NS observations (cite), and are widely considered to be candidate topologies for high-field sources \citep{Perna2014}.

\begin{figure}
\centering
  \includegraphics[width=8cm,height=7.5cm]{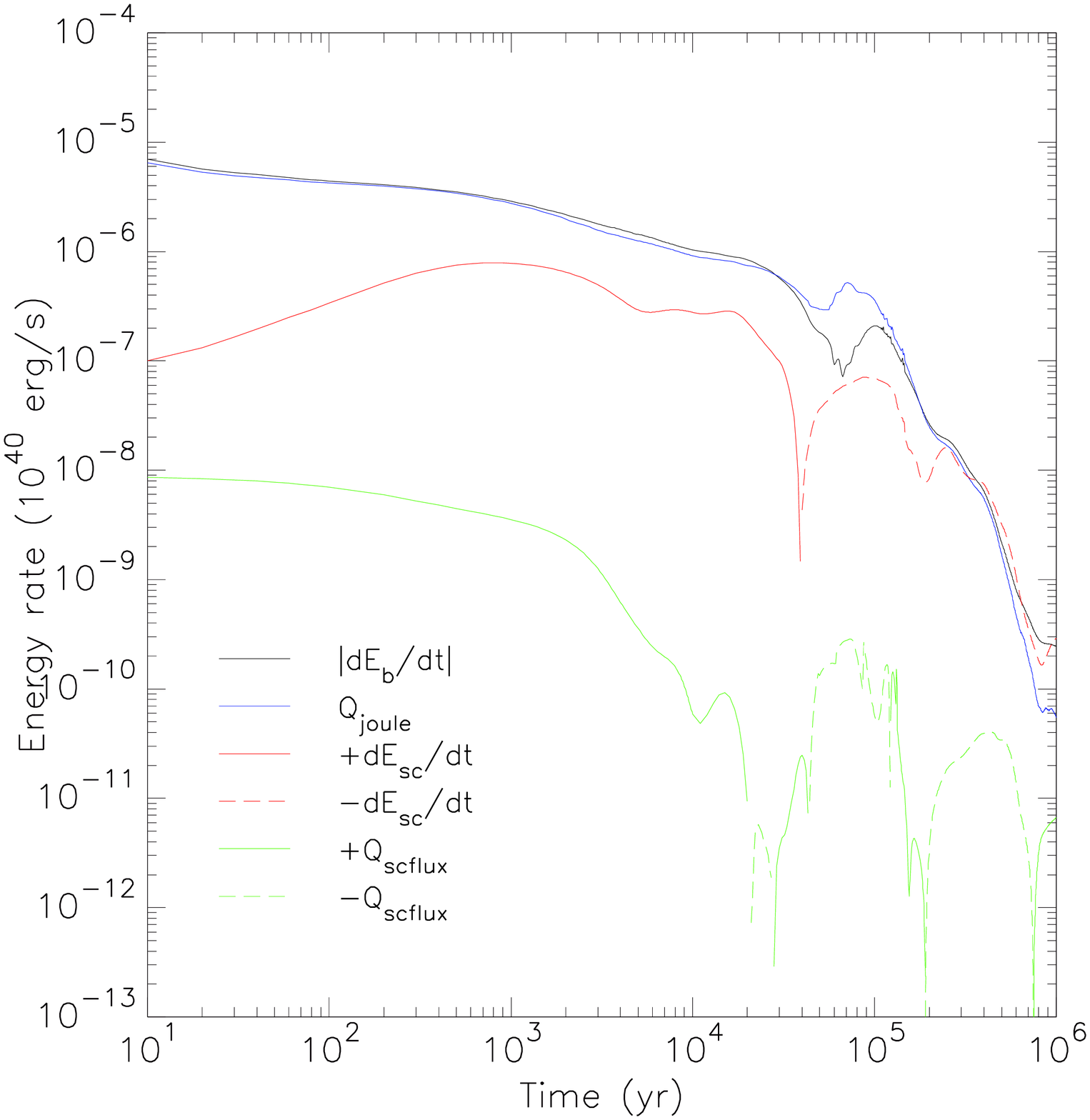}
  \caption{Energy budget in dipole case with extra crust-confined component. Total magnetic energy decay (black), superconducting magnetic energy decay/growth in core (solid/dashed red), losses from Joule dissipation (blue), radially outward/inward flux of magnetic energy at crust-core interface (solid/dashed green).}
  \label{fig:hybmixbcons}
\end{figure}

The instantaneous energy balance for this initial field prescription is illustrated in figure \ref{fig:hybmixbcons}. Most importantly, we note that the flux of magnetic energy across the crust-core interface is not uni-directional over the first megayear. Only during the first 10\,kyrs is magnetic energy expelled from the superconducting core, an effect which decreases with time. As usual, Joule dissipation dominates this early interval; the Hall cascade accelerates this process due to the intense magnetic field, as does the immediate formation of a current sheet at the crust-core interface. Field lines in the outer NS core cannot re-configure as quickly as in the crust where the fast Hall cascade pushes the field. Thus considerable field tension remains at the interface, which is released by burying the field energy into the outer NS core.

\begin{figure}
\centering
\includegraphics[width=7cm,height=6cm]{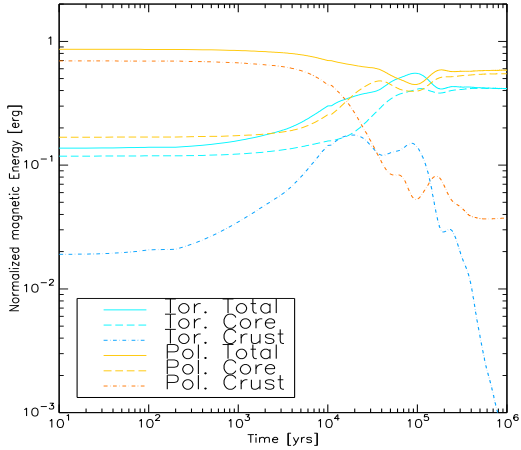}
\caption{Magnetic energy normalized to instantaneous total magnetic energy $E_b(t)$, in case of additional crust-confined field component.}
\label{fig:hybmixBenergy}
\end{figure}

Figure \ref{fig:hybmixBenergy} illustrates the evolution of poloidal and toroidal components in the core and crust, as shown before in figures \ref{fig:coreextBenergy} and \ref{fig:ffBenergy}. The energy partition favors toroidal magnetic energy after the first 10\,kyrs, due to the Hall effect acting on the strong poloidal component in the highly magnetized crust. The toroidal-poloidal coupling becomes important in the nonlinear phase (30\,kyrs - 200\,kyrs), where crustal energy is alternately transferred between poloidal and toroidal components as it is dissipated by strong currents. The slow dissipation in the core then means that most of the magnetic energy remains confined there at late times. By 1\,Myr the NS interior relaxes to a state which is indiscernible from the core-extended case. This becomes obvious when we inspect maps of the evolving magnetic field. 

\begin{figure*}
  \includegraphics[width=\textwidth]{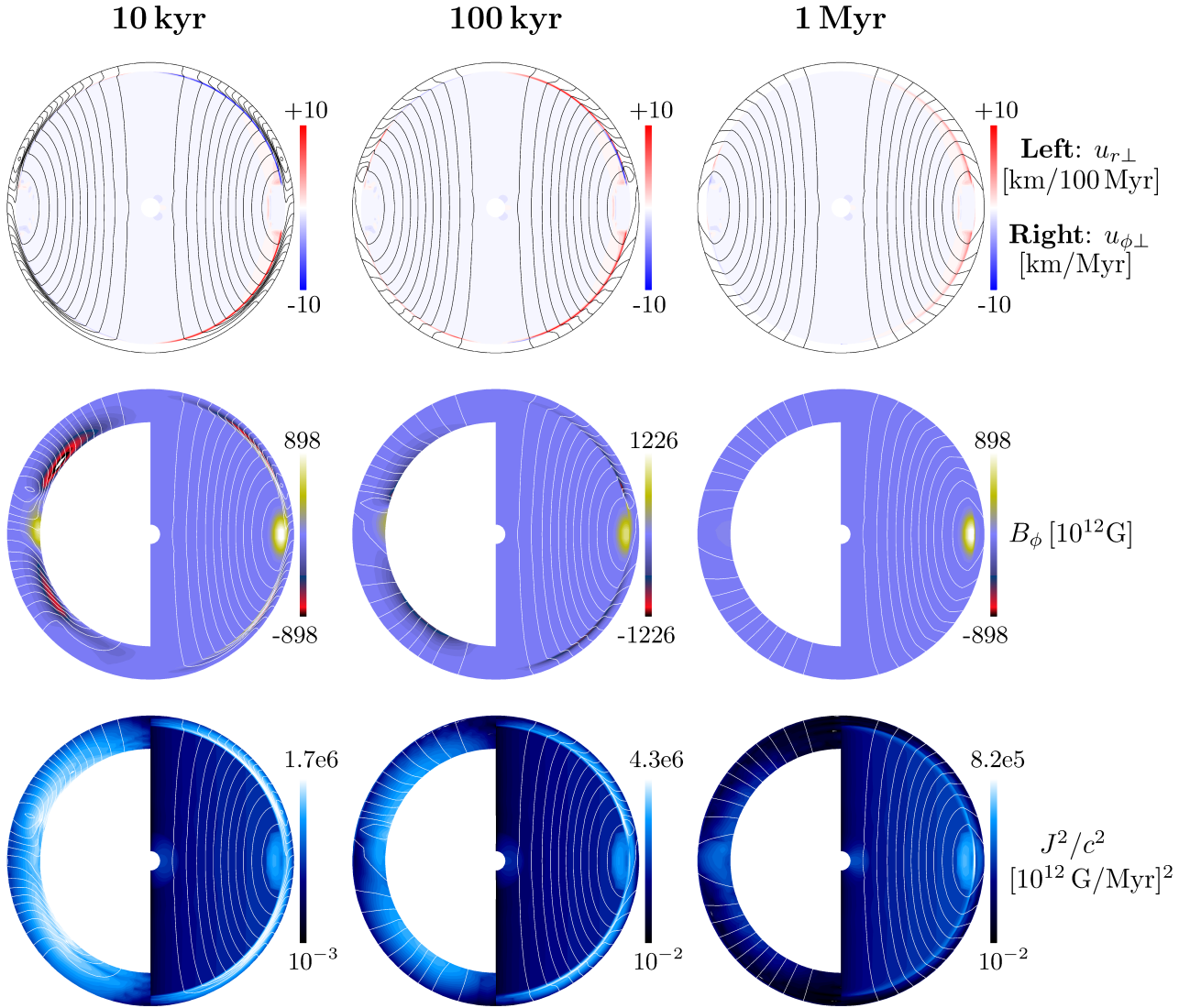}
  \caption{Snapshots of magnetic field diagnostics, extra crust-confined case. Poloidal magnetic field lines are superimposed on $u_{\bot}$ (top), $B_{\phi}$ (middle) and current intensity $J^2$ (bottom). {\bf Top row:} left half shows $u_{r\bot}$, right half shows $u_{\phi\bot}$. {\bf Middle, bottom rows:} Crust thickness exaggerated for visualization purposes in left half of each panel.}
\label{fig:hybmixbjmaps}
\end{figure*}

Figure \ref{fig:hybmixbjmaps} shows evolution and decay of the global magnetic field and currents, which are not strongly affected by the presence of the imposed toroidal component at the equator. Dynamics and cascade time-scales are again dominated by the magnetic configuration in the crust during the first 10\,kyrs, after which the characteristic kinking of the poloidal magnetic field near the crust-core interface persists. This is additional evidence that microphysical parameters at the interface could play an important dissipative role beyond 100\,kyrs. One feature to note is the absence of toroidal fluxtube drift at the equator throughout the simulation, due to the persistent toroidal field there. Recalling that drift direction is perpendicular to the magnetic field, this configuration could in principle give rise to poloidal forcing in that region, but as in \S\ref{section:initff}, the presence of a local toroidal component does not directly imply a poloidal fluxtube drift. Overall, this magnetic configuration results in evolution that is qualitatively similar to the core-extended scenario, once the crust-confined field has dissipated. However, both the azimuthal and radial drifts are faster than with the other initial conditions (see colorbar). Still, the strongest radial drift present at 100\,kyrs, which is $\sim$0.1\,km\,Myr$^{-1}$ in this case, decreases with age as the field tension at the interface is released. 

\subsection{Comments on core evolution and diagnostics}\label{section:comments}
A discussion of the direct action of this core forcing follows, based on the top panels of figures \ref{fig:cextbjmaps}, \ref{fig:ffbjmaps}, \ref{fig:hybmixbjmaps}. Upon inspection, it is immediately clear how little the field lines in the core are displaced from their initial locations. This fact appears independent of the initial conditions. Gigayear (Gyr) time-scales may be necessary to see substantial motion of inner core fluxtubes when subject to the complete set of forces that we have imposed herein, but by this age most sources would not be visible. Naturally since the inner core field remains unaffected by our imposed forcing, the poloidal drift components also remain approximately constant throughout the core volume. The exception is very close to the crust-core interface, which is the only region where the core field changes at all during the 1\,Myr simulations. In the outer core, the field lines are dragged due to the Hall-induced dynamics and decay of the crust-confined field, and the expulsion effect (or submergence) is certainly secondary to this mechanism. In all cases, fast and highly localized toroidal fluxtube drifts are immediately set up and persist for the first 100\,kyrs in interior regions where toroidal field components are not present. This localized region where the azimuthal component is confined does evolve in time in all three simulated scenarios, as the crustal field decays and experiences Hall cascade, and as the field lines change shape locally near the interface. The main result is that the core fluxtube array will continue to twist up azimuthally throughout the NS evolution. Even though the Hall effect is inactive within the liquid core, a Hall-like role is played by the differential fluxtube motion. When our proposed forces act in the NS core, a purely poloidal initial configuration of fluxtubes will reconfigure such that a toroidal field component is necessarily generated. In turn, ``microscopic currents'' ($\vec{\mathcal{J}}_{\bot}$ in equation (\ref{equation:scfull})) would flow in the core once the crust-confined field components have dissipated, meaning that Joule dissipation in the core could be substantially more effective beyond the first megayear. How this Joule-like dissipation interacts with the magnetic field present at the crust-core interface is an important question, as already suggested in \S\ref{section:simulationresults}. There, attention was drawn to oscillatory features in the energy diagnostics; specifically, radial submergence and expulsion were observed, which in turn naturally affects the energy decay or growth rate in the superconducting core. The energy oscillations have a period of $\sim$10\,kyrs - 100\,kyrs, and result from a combination of effects. Distinct descriptions of matter on either side of the interface imply the presence of strong gradients, especially in resistivity $\eta$, in thermal pressure, and in the background flow. In addition, fast Joule heating in the crust alongside slower dissipation in the core gives rise to local temperature gradients which enhance thermal diffusion. Also, there is necessarily an imbalance in the magnetic energy density because the transition from $H_{c1}$ in the core, across the interface, to $B$ in the crust may not be continuous. Thus, e.g., a parcel of crustal field with strength $B$ being submerged into the core may contribute much more than $B^2/8\pi$, since that parcel then has an energy density of $B H_{c1}/2\pi$ after the migration. In short, this critical region is susceptible to instabilities driven by local gradients, which are damped at different rates on either side of the layer. We have performed numerous additional experiments with artificially high values of the impurity factor in the deep inner crust, $Q_{\mathrm{imp}}$, in which case the oscillations are completely suppressed due to enhanced dissipation.  A more advanced understanding of the physics in this region is required to perform realistic simulations, but for now we may only say that the transport of field energy is weak, and thus the qualitative picture remains unchanged.

We have also probed the observational impact of fluxtube motion in the NS core, but refrain from including redundant figures to conserve space. Specifically, we compute solutions for spin period $P$ and period derivative $\dot{P}$ during run-time using the self-consistently computed dipolar magnetic field strength at the pole, and also compute the bolometric luminosity from the surface temperature assuming blackbody emission. We simply report that the spin-down traces are not affected by the prescribed forcing in the NS core, owing to the high core field that remains unexpelled and undissipated. The characteristic late-life knee in classic $P-\dot{P}$ diagrams is not present (see figure 10 from \citet{Vigano2013}), as the prevailing core field keeps the braking mechanism active at late times. Thus the spin-down in $P-\dot{P}$ resembles that of a constant dipolar magnetic field, i.e., straight lines, for the first megayear. The resulting families of cooling curves show no identifiable deviations from standard cooling curves based on the same initial magnetic field prescription. In short, the luminosity and observed thermal signatures exhibit no sensitivity to the imposed fluxtube drifts in the NS core. As has been reported in previous studies, the thermal evolution is primarily a function of the crustal magnetic field configuration at birth. Here this is evidenced again indirectly, in that including mechanisms which explicitly drive evolution of the core field do not alter the simulated luminosity. When we supply the extra reservoir of crust-confined energy (\S\ref{section:inithybmix}), the simulated source remains brighter during most of the visible lifetime. By about 1\,Myr, the magnetic topology resembles the core-extended dipolar case, which is reflected in the luminosity; there exists only a factor of two difference between the simulations by this time, consistent with earlier studies.

\section{Discussion}\label{section:discussion}

In this manuscript we present results from self-consistent magneto-thermal simulations which evolve the coupled NS core and crust fields, including the microphysics relevant to each region. We begin with a variety of representative initial conditions, and impose a macroscopic fluxtube drift velocity arising from an analytic treatment of fluxtube-matter interactions in the NS core. We investigate the first 1\,Myr of NS evolution, and report the following conclusions:

\begin{enumerate}

\item Expulsion of magnetic field from the NS core into the crust is not likely to affect observables within the first megayear, provided the mathematical formalism and assumptions outlined in \S\ref{section:fullforcebalance}. 

\item Simulated magnetic fields may be weakly buried in the outermost layers of the NS core, contrary to prevailing theory which predicts expulsion of the core field. 

\item Magnetic field initial conditions control the fluxtube drift patterns, and in a largely time-independent manner. The initial field prescription determines how the fluxtube array must respond (see equation (\ref{equation:vectoru3})). Because the global field is so weakly influenced by forces on the fluxtubes, there is no strong time dependence.

\item Global distortions to the core fluxtubes are confined to the outer core, where microphysics near the crust-core interface mediate dissipation and manage the magnetic energy budget, alongside the motion induced by the Hall effect in the NS crust.

\item The crust-confined magnetic field components decay on the same time-scales found in earlier works (\cite{Vigano2013} and references therein), now effectively independent of the core-crust coupling at the interface. The undissipated, remnant core field beyond $\sim 1$\,Myr is a characteristic feature that has evolutionary implications for old NSs. 

\end{enumerate}

Despite our results, expulsion of magnetic flux from superconducting NS cores could still be responsible for variations in the magnetic and thermal properties of NSs. A net expulsion does appear to occur under certain conditions during our simulations, exhibiting sensitivity to initial conditions. It also remains possible that when magnetic energy is buried into the outer core, the field could re-emerge, apparently requiring time-scales longer than a few megayears. Finally, the hypothesis that this combination of interactions in the core provides a plausible magnetar scenario appears unlikely based on the results of these simulations. After a lapse of $\sim$1\,Myr, the large-scale surface field will be dominated by the persistent field in the NS core, regardless of how the Hall-driven field evolves in the crust. It is informative to consider this global, steady-state magnetic configuration in the context of the Hall attractor reported by \citep{Gourgo2013,Gourgo2014,Wood2015}. The well-documented small scale magnetic structures driven by the Hall evolution do develop in our model--as expected--in the NS when strong crust-confined components are present, as in \S\ref{section:inithybmix}. The resulting belt of poloidal flux at the magnetic equator is consistent with earlier work \citep{Vigano2013,Wood2015}. However our initial conditions for the core field very effectively constrain the attractor state, because (a) the NS core field evolution is so slow, and (b) the global field remains anchored to the footpoints at the crust-core interface. The main result is that the Hall-Ohmic evolution of the crust-confined field is largely independent of the evolution in the NS core, and the attractor geometry is defined by the initial conditions in the core.

It remains an open challenge to understand what magnetic field morphology is present in NS interiors following crystallization of the crustal ion lattice. Modern modeling efforts suggest a highly structured, turbulent field with high sensitivity to the conditions present during the core collapse and the character of the NS progenitor. Still, unless the core magnetic field is highly concentrated in the outermost layers of the core, no enhancement of the crustal field should be detectable on megayear time-scales. Sophisticated inner boundary conditions could in principle be used to isolate the inner and outer core regions in later studies, potentially permitting designer field configurations which are particularly susceptible to the forces we impose on the core fluxtube array. For example, the development of highly tangled or turbulent microscopic fields near the crust-core interface could significantly impact the results of numerical simulations \citep{Peralta2005, Andersson2007}. This is especially true if the nuclear lattice in the inner crust is amorphous or highly disordered, which would permit a faster dissipation channel, even on microscopic length scales.

To reiterate, there are important physics which we recognize are absent from the current study. First, we exclude from our current model the computationally-expensive and poorly understood hydrodynamical evolution in the liquid core, as well as imposition of true MHD equilibrium throughout the NS interior. Thus, real NSs may have fluid degrees of freedom which are not represented in this study. In addition, our numerical model does not include effects due to rotation, nor to obliquity between the rotational and magnetic axes. The collective response of entrained fluxtubes and superfluid vortices depends on the differential rotation rates of the core and crust (Ding, Cheng \& Chau 1993). Furthermore we neglect one additional mechanism, ambipolar diffusion, which remains under serious consideration as a principle agent in driving evolution of NS cores. Lastly, we recognize that constructing and two-way coupling a representative NS magnetosphere to our numerical model may have serious physical impacts on NS simulations. 

Finally, while it has proven insightful to investigate NS evolution in this novel way, it is evident that this prescription may ultimately require a fully 3D treatment. This requires a much more sophisticated set of evolutionary algorithms, and is beyond the scope of this 2D study. However, significant advancements will be made in forthcoming publications discussing the development of such a 3D neutron star model (Elfritz et al., manuscript in preparation).

\section*{Acknowledgments}
J.G.E. and N.R. are funded by an NWO Vidi grant (PI: Rea). J.A.P. is funded by the grants AYA2013-42184-P and Prometeu/2014/6. N.R. and D.V. acknowledge support from the grants AYA2012-39303 and SGR 2014-1073. K.G. is supported by the Ram\'{o}n y Cajal Programme of the Spanish Ministerio de Ciencia e Innovaci\'{o}n. Authors also wish to acknowledge NewCompStar, COST action MP1304 for partial funding of this research. We also thank the anonymous referee for constructive suggestions.

\label{lastpage}

\end{document}